\documentclass[aps, prb, groupedaddress, showpacs, showkeys, twocolumn, 
]{revtex4}
\usepackage{graphics, amsfonts, amsmath, mathrsfs, amssymb, hyperref, manfnt}
\def\g{\mbox{\sl g}}
\begin{document}
\title{Nonequivalence of equivalence principles}

\author{Eolo Di Casola}
\email{\tt eolo.dicasola@sissa.it} 
\author{Stefano Liberati}
\email{\tt stefano.liberati@sissa.it} 
\affiliation{SISSA, Via Bonomea 265, 34136 Trieste, Italy}
\affiliation{INFN, Sezione di Trieste, Via Valerio 2, 34127 Trieste, Italy}

\author{Sebastiano Sonego}
\email{\tt sebastiano.sonego@uniud.it}
\affiliation{DCFA, Sezione di Fisica e Matematica, Universit\`a di Udine, Via delle Scienze 206, 33100 Udine, Italy}

\begin{abstract}
Equivalence principles played a central role in the development of general relativity.  Furthermore, they have provided operative procedures for testing the validity of general relativity, or constraining competing theories of gravitation.  This has led to a flourishing of different, and inequivalent,  formulations of these principles, with the undesired consequence that often the same name, ``equivalence principle'', is associated with statements having a quite different physical meaning.  In this paper we provide a precise formulation of the several incarnations of the equivalence principle, clarifying their uses and reciprocal relations.  We also discuss their possible role as selecting principles in the design and classification of viable theories of gravitation.
\end{abstract}

\thispagestyle{empty}
\maketitle


\section{Introduction}
\label{S:1}

Virtually every textbook on general relativity contains, at some stage, a discussion of the principle of equivalence.\cite{synge}  However, the formulations of the principle that one finds in the various textbooks differ.  The problem arises, then, as to whether these formulations are equivalent.

In fact, they are not, and the most scrupulous authors distinguish between ``weaker'' and ``stronger'' versions.  Nevertheless, it often remains unclear to the student which are exactly the implications of each formulation, both in terms of physical content and mathematical structure~\cite{bk}.  The goal of the present paper is to contribute to a clarification of these issues.

Apart from reasons related to the pedagogy of general relativity, however, there is also another motivation for this effort.  It might be useful to be able to classify the ceaselessly growing collection of alternative theories of gravity  not only in terms of their mathematical structure but also using simple statements about the general behaviour of physical systems, thus providing a possible connection with experiments.  This is the way in which the equivalence principle has been used {\em de facto\/} in the last half century.  Each of the various inequivalent formulations of the principle identifies a different class of theories, with different fundamental mathematical content.  Of course, it is clear that this programme demands each version of the equivalence principle to be formulated unambiguously (see, {\em e.g.},~Ref.~\onlinecite{Sotiriou:2007zu} for an extended discussion on this issue).  

The plan of the paper is the following.  We first present, in Sec.~\ref{S:2}, the hierarchy of principles, from the simplest (Newton's equivalence principle) to the most complex (the strong equivalence principle).  Then, in Sec.~\ref{S:3}, we discuss the logical relationships among such statements.  In Sec.~\ref{S:4} we  spell out the structural role of equivalence principles within gravitational theories.  Section~\ref{S:5} contains some comments about Einstein's original formulation of the principle, and about future perspectives of a possible renaissance of the equivalence principles. Indeed, the vast and intricate landscape of modified and extended theories of gravity emerged so far under the pressure of the latest wealth of cosmological data strongly calls for new, effective selection criteria, among which the unforgotten equivalence principles may play a key role.

\section{Equivalence principles}
\label{S:2}

We want to give an operational formulation of the various versions of the principle of equivalence --- a formulation that makes reference only to the behaviour of physical systems, without involving theoretical paraphernalia like the notion of a metric, a connection, etc.  The reason behind this choice is that we do not want to select general relativity from the outset; we wish instead to keep our statements as general as possible, so that we can cover at once theories that use different mathematical structures.  The only exception to this rule concerns spacetime itself, which we regard as a smooth manifold $\mathscr{M}$ (usually taken as four-dimensional), equipped with some notion of past and future.  Physical fields are geometric (tensorial or spinorial) objects over $\mathscr{M}$.  This framework includes, of course, all theories modelled after general relativity, but is wide enough to cover also, e.g., Newton's theory of gravity, reformulated {\em \`a la\/} Cartan.\cite{malament}  An observer is idealised as a future-directed smooth causal curve on $\mathscr{M}$, and a reference frame in some spacetime region $\mathscr{U}\subseteq\mathscr{M}$ is a family of observers, such that each event in $\mathscr{U}$ belongs to one and only one of them.

A preliminary warning: there is, unfortunately, no general agreement about the terminology.  For example, our ``Einstein's equivalence principle'' (Sec.~\ref{Ss:EEP} below) is sometimes called the ``weak equivalence principle'', and sometimes the ``strong equivalence principle''.  This confusion is more common in the older literature, but still persists nowadays in spite of some attempts at establishing a common glossary.\cite{tll}  We basically adhere to the nomenclature presented in Will's influential book,~\cite{will} which follows the proposal in Ref.~\onlinecite{tll}.  The specific formulation of the various principles is, however, rephrased to suit the logic of our discussion.

\subsection{Newton's equivalence principle}
\label{Ss:NEP}

All viable theories of gravity must reduce, in some limit, to Newton's theory,\cite{TeVeS} so we begin recalling the basic elements of the latter.  The Newtonian gravitational potential $\Phi(\boldsymbol{x},t)$ obeys Poisson's equation
\begin{equation}
\nabla^2\Phi=4\pi G\rho\;,
\end{equation}
where $G$ is Newton's constant, and $\rho(\boldsymbol{x},t)$ is the density of active gravitational mass (the property of matter playing the role of  ``source'' for gravity, analogous to the ``active electric charge'' that generates an electric field).  The equation of motion for a particle\cite{fn3} with inertial mass $m_\mathrm{i}$ in a gravitational field with potential $\Phi$ is 
\begin{equation}
m_\mathrm{i}\,\ddot{\boldsymbol{x}}(t)=-m_{\mathrm{g}}\boldsymbol{\nabla}\Phi(\boldsymbol{x}(t),t)\;,
\end{equation}
where $\boldsymbol{x}(t)$ is the particle position at time $t$, and $m_{\mathrm{g}}$ is the particle's passive gravitational mass (the property that expresses how a particle ``feels'' a given gravitational field, analogous to the ``passive electric charge'' one has inside the expression of the Lorentz force).  

In principle, the active and passive gravitational masses express different properties, and might well be distinct.  If one requires, however, that the action-reaction principle holds in the Newtonian limit, it is easy to see that they must coincide up to an irrelevant universal coefficient.~\cite{berry}  Thus, from now on we shall identify them, and speak only of a ``gravitational mass'' $m_{\mathrm{g}}$.  

On the other hand, the basic principles of dynamics do not say anything about possible relations between the gravitational mass $m_\mathrm{g}$ and the inertial mass $m_\mathrm{i}$.  Indeed, these quantities express, in principle, very different properties of a body.  Experiments, however, show that they are also proportional to each other, through a universal coefficient, with great accuracy, and can therefore be made equal by a suitable choice of units.  Loosely speaking, this means that every form of matter-energy responds to gravity in the same way.  This remarkable fact was established by Newton,\cite{hertz} who observed that simple pendula whose bobs are made of different materials and have different weights, but all having the same length, oscillate with the same period.\cite{newton}  (Strictly speaking, this method tests the equality between $m_\mathrm{i}$ and $m_\mathrm{g}$ only indirectly, as a consequence of the universality of free-fall within Newtonian dynamics.  However, that $m_\mathrm{i}=m_\mathrm{g}$ can be verified independently, though much less accurately; see Sec.~\ref{S:3}.)  It is therefore appropriate to call this empirical fact

\medskip

\noindent{\bf Newton's  equivalence principle (NEP):} {\em In the Newtonian limit, the inertial and gravitational masses of a body are equal.}

\medskip

Note that NEP is formulated in such a way that its validity can be tested for theories other than Newton's.  The requirement that one works in the Newtonian limit is because only in those conditions one can unambiguously identify, in general, an inertial and a gravitational mass.  Indeed, $m_\mathrm{i}$ and $m_\mathrm{g}$ are eminently Newtonian quantities, whose definition might turn out to be problematic in other theories.

\subsection{Weak equivalence principle}
\label{Ss:WEP}

The next version of the equivalence principle is nothing but the old, empiric law of universality of free fall --- a lively relic of Galilei's grand achievements.\cite{uff}  The statement reads:

\medskip

\noindent{\bf Weak equivalence principle (WEP):} {\em Test particles with negligible self-gravity behave, in a gravitational field, independently of their properties.}

\medskip

It is understood that the particles are not subject to any force of non-gravitational origin and that ``behave independently of their properties'' means that the future histories of the particles will be the same, provided they have the same initial conditions.  Of course, it is important that the notion of ``test particles with negligible self-gravity'', used in the statement, be absolutely clear and unambiguous.  By a ``test particle'', we mean one which does not back-react on the surrounding environment (although it is acted upon by the environment itself, of course).  This notion is clearly valid only within some approximation.  

Establishing what ``negligible self-gravity'' means is a bit more tricky.  For this purpose, it is convenient to define the dimensionless parameter 
\begin{equation}
\sigma:=\frac{Gm}{c^2 r}\;,
\end{equation}
where $m$ denotes the mass of the body (it does not matter whether inertial or gravitational, if the two masses are very close to each other), and $r$ is a convenient measure of its size.\cite{fn4}  The parameter $\sigma$ expresses the ratio between the gravitational energy of the body (of order $Gm_\mathrm{g}^2/r$) and its rest energy $m_\mathrm{i}c^2$, so it is a good indicator of the presence and amount of self-gravity.  For pebbles, planets, or even a star like the Sun, $\sigma$ is very small, whereas it becomes of order $1$ for compact objects such as neutron stars or black holes.  Indeed, since $\sigma$ is of the same order of magnitude as the ratio between the Schwarzschild radius of a body and its size $r$, it can also be regarded as an indicator of the compactness of the body.

Note that the two notions of ``test particle'' and ``negligible self-gravity'' are logically independent.  A pebble is, with excellent approximation, a test particle in the gravitational field of the Earth, and its self-gravity is negligible.  However, the Moon has $\sigma\ll 1$, but for several purposes cannot be regarded as a test body, as it affects the gravitational field of the Earth in a non-negligible way.  And a micro black hole (i.e., one having a very small mass) might well be considered a test particle, although its parameter $\sigma$ attains the maximum possible value, regardless of $m$.

\subsection{Gravitational weak equivalence principle}
\label{Ss:GWEP}

Removing the condition that self-gravity be negligible, one obtains the

\medskip

\noindent{\bf Gravitational weak equivalence principle (GWEP):} {\em Test particles behave, in a gravitational field and in vacuum, independently of their properties.}

\medskip

Of course, the GWEP includes the WEP, which corresponds to the limit $\sigma\to 0$.  Note that the condition that we are dealing with a test particle has not been relaxed.  The reason is that if one were also considering bodies whose back-reaction on the environment is non-negligible, it would be impossible to compare their behaviour and then establish whether it is, or is not, the same.  In other words, the very notion of a universal behaviour requires that the environment is unchanged when different particles are considered, and this can happen only if they are test particles.

On the other hand, the formulation of the GWEP contains the explicit restriction to a vacuum (that is, an empty background), which was not present in the WEP. The reason is that, since now we are considering particles whose own gravitational field is non-negligible, they would exert some force on whatever surrounding matter might be present. By the action-reaction principle, there would then be a force on the particle as well, which would make its world-line deviate from the one of a particle with no self-gravity, thus undermining universality. Removing the restriction to vacuum, thus, no theory would satisfy the GWEP.\cite{edc}

\subsection{Einstein's equivalence principle}
\label{Ss:EEP}

The principles presented so far all refer only to the mechanical behaviour of particles.  The following statement covers instead all non-gravitational physics.

\medskip

\noindent{\bf Einstein's equivalence principle (EEP):} {\em Fundamental non-gravitational test physics is not affected, locally and at any point of spacetime, by the presence of a gravitational field.}

\medskip
 
Again, a few words of explanation are in order.  By ``non-gravitational test physics'' we mean, of course, processes involving the mechanics of particles and continua, thermodynamics, electromagnetism, etc., provided that they do not affect a pre-existing background gravitational field (or do not create a non-negligible one, in case there is none).  Thus, particles, fluids, fields, and so on, are all considered ``test entities'' from the gravitational point of view.  

When we say that a phenomenon is ``not affected, locally, by the presence of a gravitational field'', we mean the following.  Imagine to set up an experiment  taking place in a sufficiently small spacetime region $\mathscr{U}$, where a gravitational field exists, and to record the results.  Then, it is always possible to choose a suitable reference frame, in a region $\mathscr{U}'$ where gravity is absent, whose observers will find the same results when performing the experiment following the same procedure. In most textbooks, EEP is formulated in terms of the equivalence between a locally non-rotating, freely-falling frame in a gravitational field and an inertial frame in the absence of gravity. This is, probably, the simplest situation, and it is fine to use it for pedagogical purposes. It should be clear, however, that other choices are possible (for instance, choosing a suitably accelerated frame in the absence of gravity to simulate a gravitational field), and that the general idea at the core of EEP is the correspondence between local frames in a gravitational field and frames in the absence of gravity. 

The statement of EEP, and the explanation provided, contain the slippery terms ``locally'' and ``sufficiently small''.  The catch is that, in fact, physical processes in a gravitational field do unfold differently than in the absence of gravity.  Sometimes, it is claimed that any difference can be detected only by experiments performed over finite scales of distance and time, and that such differences become smaller and smaller as the size of $\mathscr{U}$ decreases.  This would suffice to justify a condition about locality in the statement of the principle.  The claim as it stands, however, is just plainly wrong.  For example, consider an experiment where the distance $\ell$ between two nearby freely-falling particles is measured, and consider the quantity $\ddot{\ell} / \ell$, where a dot denotes the derivative with respect to proper time, or to a convenient laboratory time.  In an inhomogeneous gravitational field, this quantity does not vanish, in general, even in the limit $\ell\to 0$, although it does vanish identically in the absence of a gravitational field.  Indeed, such quantity is proportional to a component of the tidal tensor $\partial^2\Phi/\partial x^i\,\partial x^j$, which is obviously a local object characterising the gravitational field.\cite{tides}  Other examples of experiments performed within arbitrarily small spacetime regions, whose result depends on whether there is, or there is not, a gravitational field, have been provided in Ref.~\onlinecite{ohanian}.  It is easy to realise, however, that in all such cases one is dealing with the behaviour of systems that are not elementary, but composite.  Hence, the need to restrict the statement of EEP to ``fundamental physics''. The latter is evidently an unsatisfactory requirement, given our ignorance of the basic laws. However, in many situations, it is clear at least which physical laws should {\em not\/} be regarded as fundamental (so one does not expect that EEP necessarily holds for them). For instance, in the context of point particle mechanics, this is the case for all those types of ``effective'', ``reduced'' equations trying to describe the average behaviour of a composite system (e.g., the motion of a centroid). Such equations can in fact be derived from the equations of motion for the single particles --- for which EEP {\em does\/} hold --- together with the knowledge of the interactions between them. Similarly, particles with spin or multipole moments do not follow geode- sics and violate the WEP, hence also EEP, because their equations of motion contain terms involving the curvature.\cite{spin} These violations, however, only follow from the fact that such equations arise taking the point particle limit of extended bodies with an internal structure. As a final example, one can consider the phenomenon of geodesic deviation. As already noted, this violates EEP, although the basic ingredient used to derive it --- the geodesic equation --- does not.

The statement above of EEP is general, and is compatible, in principle, with any spacetime structure.  To the best of our knowledge, however, the laws of physics in the absence of gravity are Poincar\'e-invariant.  This means that they are invariant both under spacetime translations, and under the action of the Lorentz group (containing spatial rotations and arbitrary boosts).\cite{LSB}  For this reason, EEP is often stated within this more specific context, as the joint requirement of the WEP, the ``local position invariance'',  and the ``local Lorentz invariance'' for non-gravitational test experiments.\cite{will}

\subsection{Strong equivalence principle}
\label{Ss:SEP}

Just as removing from the WEP the condition that self-gravity be negligible leads us to the GWEP, we can extend EEP to include also gravitational phenomena.

\medskip

\noindent{\bf Strong  equivalence principle (SEP):} {\em All test fundamental physics (including gravitational physics) is not affected, locally, by the presence of a gravitational field.}

\medskip

At first, this statement sounds odd. How can gravitational physics not be affected by the presence of a gravitational field?  However, as for EEP, the physical processes considered here must be regarded as {\em test experiments\/} over a background that is not significantly affected by them.  Thus, the kind of gravitational experiments to which the SEP applies can be, for example, the mutual attraction between two sufficiently light bodies, or the detection of a weak gravitational wave.  The principle then says that, even if such experiments are performed within some background gravitational field, it is always possible to find observers who will (locally) record the same results, when performing the same experiments in the absence of such a background field.  This is by no means obvious or granted, in a non-linear theory of gravity as general relativity, where distant masses could have a non-trivial  effect on local gravitational processes.\cite{bertotti}

As for EEP, also for the SEP it is common to find, in the literature, a formulation that makes directly reference to the Minkowskian structure of spacetime in the absence of gravity.  Thus, the SEP is implemented  requiring local position invariance and local Lorentz invariance for all test experiments (including those involving gravity), together with the validity of the GWEP.\cite{will}

\section{Relationship between the various formulations}
\label{S:3}

Obviously, the various principles introduced in the previous section are not all equivalent.  We now analyse their mutual relationships.

It is easy to realise that the WEP implies NEP.  For, if NEP were false, then the universality of free-fall would not hold even in the Newtonian approximation, so the WEP would also be false.  In fact, the two principles are often identified, as if the validity of NEP guaranteed that also the WEP hold.  This is, however, not true.  Only as long as the two kinds of masses enter in the equations of motion through their ratio $m_\mathrm{i}/m_\mathrm{g}$ alone, as it happens in Newton's theory, does NEP imply the universality of free-fall.  If different sorts of combinations of $m_\mathrm{i}$ and $m_\mathrm{g}$ are allowed, then spurious instances of the masses crop up and generically the WEP fails.\cite{fn5}  It should be noted, in this respect, that although NEP is usually tested through the WEP (see, e.g., Newton's experiments with pendula, already mentioned in Sec.~\ref{Ss:NEP}), the  proportionality between $m_\mathrm{i}$ and $m_\mathrm{g}$ can be established making use of procedures not involving the WEP.  For example, $m_\mathrm{i}$ can be measured applying a known non-gravitational force to the body, and measuring its acceleration; $m_\mathrm{g}$ can be measured weighing the body on a spring scale.  

In an empty background, the GWEP implies NEP for self-gravitating bodies.  Conversely, if NEP fails for self-gravitating bodies, then the GWEP is violated.  This is indeed a rather common subject of investigation.  Given a gravitational theory in which the WEP holds, it is possible to verify whether NEP holds as well for a body with non-negligible self-gravity.  For this purpose, it is appropriate to distinguish NEP from its extension to self-gravitating bodies --- we shall call the latter  Gravitational NEP (GNEP).  Calculations show that, although in general relativity $m_\mathrm{i}=m_\mathrm{g}$ always, this is a rather exceptional situation (see Ref.~\onlinecite{nordtvedt} for the failure of the GNEP in Brans--Dicke theory).  An interpretation of this result is that, contrary to what happens in general relativity, in most of the theories gravitational energy does not respond to gravity as the other forms of energy do.  It is worth noticing, in this respect, that experiments such as those involving the laser ranging of a self-gravitating body like the Moon, which measure the dependence of the body’s orbit on the ratio $m_\mathrm{g}/m_\mathrm{i}$, actually test the GNEP, rather than the GWEP.

One easily realises, also, that EEP implies the WEP.  Indeed, EEP guarantees that the behaviour of a freely-falling particle in a gravitational field is, locally, indistinguishable from that of a free particle in the absence of gravity (in the sense, discussed in~\ref{Ss:EEP}, that there are two reference frames whose observers see the same type of motion for the particles).  Since the behaviour of free particles is universal, it follows that also freely-falling particles behave in a universal way.\cite{noWEP}  By the same token, the SEP implies the GWEP (and of course EEP).

Whether the WEP implies EEP is still an open issue, usually referred to as  Schiff's conjecture\cite{tll, will}. This claim has never been proved in full generalitybut only within some simple model contexts, such as composite bodies bound via classical electrodynamics in a spherically symmetric gravita- tional background.\cite{will, lightman-lee}  If the conjecture were correct, one could also establish a connection between the GWEP and EEP, using the implication ``GWEP $\Rightarrow$ WEP'' as a first step. 

A rationale behind Schiff's conjecture, is that ``test particles'' are ultimately compound objects, held together by forces of various type.  It is not obvious that, if such forces were sensitive to the presence of a gravitational field (that is, if EEP were false), a universal behaviour for ``test particles'' could emerge.   If this were indeed the case, a violation of EEP would entail a violation of the WEP, which would be logically equivalent to the statement ``WEP $\Rightarrow$ EEP''.

In spite of this circumstantial evidence, however, one should keep in mind that a universal emergent behaviour, independent of the details of the underlying microphysics, is what one often observes for macroscopic systems, and that something similar might happen for the mechanical behaviour of test bodies.  Which means that a sound proof of the conjecture is definitely needed before any definitive statement could be drawn.

As we said, the SEP is just EEP extended to test gravitational phenomena, of which the free-fall of a test body with self-gravity is only a particular case.  An interesting problem is then whether adding the GWEP to EEP one recovers the full SEP.  To our knowledge, this ``gravitational version'' of Schiff's conjecture has never been formulated before.  If correct, it would establish the GWEP as the key element of the SEP, the one distinguishing the latter from EEP.  Note that the conjecture is {\em not\/} that the GWEP implies the SEP, but that the SEP is equivalent to the union of the GWEP {\em and\/} EEP.  To better understand this point, one might consider, for example, the propagation of a weak gravitational wave.  Of course, the GWEP has nothing to say about such phenomenon, which does not involve any gravitational self-interaction, and thus cannot be used to infer what the propagation on a curved spacetime is, knowing the one on a flat background.  Still, a weak gravitational wave is equivalent to a spin-2 field, to which EEP can be applied.

\section{Theoretical role of the equivalence principles}
\label{S:4}

In this section, we elucidate how the equivalence principles enter in the construction of a theory of gravity, in terms of the mathematical structure they suggest.

The role of the WEP in theory building is to hint that the response to a gravitational field can be described by an affine connection on spacetime --- that is, by a rule for parallel-transporting a vector along a path on $\mathscr{M}$.  For later convenience, we denote such a mathematical structure by $\Gamma$.  The logic behind this daring association goes as follows.  Universality of free-fall indicates that the worldlines of test particles in a gravitational field do not depend on the particle properties, but only on their gravitational environment.  Therefore, a gravitational field is associated with a set of preferred lines in spacetime, which can be considered the autoparallel lines of some connection $\Gamma$ (i.e., those lines whose tangent vector, parallel-transported along the line itself, remains tangent).\cite{projective}  When there is no gravitational field, the preferred worldlines reduce to the straight-line inertial motions of Newtonian mechanics and special relativity, which correspond to a fixed, flat connection $\Gamma_0$.  Conversely, in the presence of a gravitational field, the connection is, in general, dynamical and curved, as  implied by elementary considerations about tidal forces.

It is perhaps worth remarking that identifying gravity with spacetime geometry is not a compulsory step, but follows from  adopting a view in which all universal features are ascribed to geometry.\cite{reichenbach}  According to this philosophy, the universal behaviour of freely-falling test particles is better explained by spacetime geometry than by a physical field (gravity) that couples to all bodies in the same way.\cite{fn7}

At this stage, we still have wide freedom in constructing a theory of gravity.  In the so-called {\em metric theories\/} (of which general relativity is an example), one assumes that, since in the absence of gravity spacetime possesses a Minkowski metric $\eta$ --- of which $\Gamma_0$ is the Levi--Civita connection --- the same link holds when a gravitational field is present.   Thus, the non-flat $\Gamma$ is regarded as the Levi-Civita connection of a spacetime metric $\g$, different from $\eta$ and dependent on the matter environment.  The WEP alone, however, is not enough to justify these further steps. For instance, Newton-Cartan's theory includes the WEP, yet it is based on a non-metric affine connection --- in fact, there is no spacetime metric in Newton-Cartan's theory.

The role of EEP is exactly to fill in this gap.  Together with the independent  information that, in the absence of gravity, the physical laws fit within the framework of special relativity, it tells us that the structure of the spacetime associated with a gravitational field must be locally Minkowskian.  In other words, the curved connection $\Gamma$ must be locally indistinguishable  from the flat Levi--Civita connection of the Minkowski metric $\eta$.  (Of course, what ``locally'' exactly means depends both on the curvature scale and on the degree of accuracy required.)  This is possible if and only if $\Gamma$ itself is the Levi-Civita connection of some metric $\g$.

However, EEP allows us to go farther than this.  That $\Gamma$ is the Levi-Civita connection of some metric, could be derived from the simpler requirement that local {\em chronogeometric\/} measurements --- measurements that can ultimately be reduced to those of spatial distances and time intervals --- do not reveal the presence of a gravitational field.  But EEP says more; namely, that this must be the case for all non-gravitational fundamental physical phenomena, i.e., that fundamental non-gravitational physics in a curved spacetime is locally Minkowskian.  Therefore, not only does EEP select metric theories of gravity; it also provides us a powerful prescription for writing the various physical laws in a curved spacetime, {\em viz.}, for coupling gravity to the other fundamental physical phenomena.\cite{sw}  

It is worth pointing out that, sometimes, this prescription is misleadingly understood as a rule for ``minimal coupling'', the idea that the basic equations of physics in the presence of gravity differ from their counterparts in flat spacetime only by the replacements of the Minkowski metric $\eta$ by a curved one $\g$ and, correspondingly, of the partial derivatives by the covariant derivatives associated with the Levi-Civita connection of $\g$. This is, however, not always correct.  Strictly speaking, what EEP requires is that, under given conditions, physical phenomena in a sufficiently small region of spacetime unfold in the same way when there is, or is not, a gravitational field. This is clearly a condition on {\em solutions}, rather than on {\em equations}. Since any solution can be constructed in terms of an elementary one, called the ``Green function'', such a condition can be implemented, mathematically, by requiring that the local structure of the Green function associated with a physical law be the same in a curved and in a flat spacetime. That the prescription on the Green function is not equivalent to the one on the equations, and that it is the former that is correct, can be seen by analysing the case of a scalar field.\cite{sv}  Minimal coupling can lead to all sort of local bizarre behaviour (massive fields propagating along the light-cone, massless fields propagating inside it); this would allow an experimenter to tell the presence of a gravitational field working on arbitrarily small regions, in manifest violation of EEP.  On the contrary, the requirement on the Green function automatically guarantees that any solution behaves, locally, as if spacetime were flat.

Summarising: both the WEP and EEP hint at the mathematical structure of spacetime in the presence of gravity.  Moreover, EEP can be turned into a prescription for writing down physical laws in a gravitational field, once they are known in Minkowski spacetime.  What do then the GWEP and the SEP say, in addition to this all?

Among the metric theories of gravity selected by EEP, one can further distinguish the so-called {\em purely metric\/} theories, in which gravity is exclusively described by the metric $\g$ alone. Thus, non-purely metric theories contain other types of gravitational fields, in addition to $\g$. This is the case, for instance, for the Brans-Dicke theory, where gravity is described by the metric and by a scalar field.

It turns out that the GWEP is satisfied only by purely met-ric theories. A detailed formal proof of this statement can be found in Ref.~\onlinecite{edc}. Physically, what happens is the following. In a non-purely metric theory, the mass-energy of a self-gravitating body acquires a dependence on the extra gravitational fields (e.g., the scalar field for the Brans-Dicke theory). In a nontrivial background, where such extra fields are not constant, this dependence produces a force, which makes the motion of the body non-geodesic.

It is worth mentioning that this restriction is highly nontrivial, because in four dimensions it pins down just Einstein’s theory (with an arbitrary cosmological constant). Indeed, other seemingly purely metric theories of gravity, whose Lagrangian contains higher powers of curvature --- in contrast to the Einstein-Hilbert one, which is just the Ricci scalar --- are actually scalar-tensor theories in disguise.\cite{edc} Since Einstein’s theory is the only one known to satisfy the SEP, it then seems that the “gravitational Schiff conjecture” formulated at the end of Sec.~\ref{S:3} is correct.

\section{Concluding remarks}
\label{S:5}

We have seen that the various formulations of the equivalence principle form a hierarchy (or rather, a nested sequence of statements narrowing down the type of gravitational theory), from the simple NEP to the sophisticated SEP.

It is worth adding that, within such a hierarchy, the principles can also be broadly separated into two classes.  Some of them (NEP, WEP, and GWEP) are just statements about properties and the behaviour of particular physical systems.  The others (EEP and SEP) can be reformulated as general  ``impossibility principles''.  More specifically, EEP and the SEP forbid the detection of a gravitational field by means of local experiments (albeit restricted to those of a fundamental type).  In this sense, they are very similar to the relativity principle, which forbids the detection of one's state of motion by internal experiments within the class of inertial frames.  

Remarkably, this was Einstein's original view of the equivalence principle.\cite{norton}  He regarded the indistinguishability, through internal experiments, between an accelerated laboratory in Minkowski spacetime and a laboratory at rest in a suitable gravitational field, on a similar footing as the principle of special relativity.  He thought that, just as the latter establishes the relative character of velocity (hence, the equivalence between all inertial frames), so the equivalence principle establishes that also acceleration is a relative quantity.  With this interpretation, the equivalence principle generalises the principle of relativity from the class of inertial frames to frames in arbitrary motion --- indeed, this is the origin of the name ``{\em general\/} relativity'' for Einstein's theory of gravity.

A caveat is in order, at this point.  To maintain Einstein's view, the equivalence between an accelerated laboratory in the absence of gravity, and a laboratory at rest in a gravitational field (or between an inertial laboratory and a freely-falling one) should be {\em exact\/}.  This is not the case, for several reasons.  As we saw in Sec.~\ref{Ss:EEP}, the presence of curvature requires one to restrict attention to sufficiently small regions (infinitely small, in principle) and to phenomena that are sufficiently elementary; otherwise the difference between the two situations can be detected.  Clearly, adding these conditions weakens considerably the interpretation of EEP and the SEP as  generalised principles of relativity, which is probably why their role in the presentations of the theory changed over the years.  To be fair, Einstein never used this ``punctual'' version of the principle, and used to speak of the equivalence between a uniformly accelerated laboratory in Minkowski spacetime and a laboratory supported within a uniform gravitational field.  Yet, since the latter concept is ill-defined even within the context of general relativity itself,\cite{munoz} this line of reasoning is nowadays somewhat discredited, except as an approximate heuristic tool.

There is however, at the present time, a new reason  for going back to the equivalence principles.  The puzzling picture of the universe emerging from the latest cosmological data~\cite{Ade:2013zuv} has in the last years propelled a renewed interest in alternative theories of gravity, leading to a whole new blossoming of novel ideas and suggestions.  Furthermore, the surprising duality between gravitation in an anti-de Sitter bulk spacetime in five dimensions, and a conformal field theory on the four-dimensional boundary of the AdS bulk\cite{AdS} has motivated studies of gravitation in higher dimensions, where general relativity is no longer the only non-trivial, purely metric, theory of gravity with second-order field equations --- actually, in more than four dimensions there is a whole hierarchy of actions leading to at most second-order field equations for the metric, the so-called Lanczos--Lovelock actions.\cite{Padmanabhan:2010zzb}  This new complexity seems to require one or more ``ordering principles" to classify  theories of gravity,  somehow identifying what makes general relativity ``special'' (apart from simplicity, of course).  The GWEP as described in this paper seems to provide a reasonable starting point in this sense, because it selects exactly the Lanczos-Lovelock theories appropriate to the given spacetime dimensionality.\cite{edc}


The fact that the GWEP appears so selective suggests a possible interpretation of it in terms of the nonlinearity of the gravitational field equations. Although the WEP is often postulated for the motion of particles in a gravitational field (one assumes that the world-lines of freely-falling test par- ticles are geodesics), in some theories one can actually {\em derive\/} the equations of motion for small self-gravitating bodies from the gravitational field equations alone.\cite{infeldschild, EIH} The logic is then the following: the field equations for gravity dictate the motion of small self-gravitating bodies, which includes also the motion of particles with negligible self- gravity as a limiting case. In this sense, the GWEP (and, of course, the WEP) can be regarded as a constraint on the field equations themselves. In such theories, there is no freedom to postulate the equations of motion, even for test particles without self-gravity. This is the case, for instance, in general relativity.\cite{infeldschild, EIH} Not all gravitational theories, however, are powerful enough to do so. Basically, the equations of motion for a small body follow from the gravitational field equations if the latter imply local energy-momentum conservation, which amounts to four equations. For this to happen, the theory should be invariant under arbitrary coordinate transformations (the so-called {\em diffeomorphism invariance})— a requirement equivalent to asking that it does not contain non-dynamical quantities.\cite{conserv} This condition, in turn, implies that the equations for the gravitational field must be sufficiently ``complicated''.\cite{fn8} Moreover, since for a self-gravitating body the local conservation of energy-momentum must also include the energy-momentum content of the body’s own gravitational field, no linear field equation can do the job. Hence, if the GWEP has to be a consequence of the gravitational field equations, the theory of gravity must be nonlinear.\cite{weatherall}

Actually, one can say more. The SEP implies that, for test gravitational experiments, the gravitational background is locally irrelevant --- the outcome is indistinguishable from that of an identical experiment performed in Minkowski spacetime. This is not what one would expect in an arbitrary nonlinear theory.\cite{bertotti} Thus, the GWEP and the SEP play the role of selection rules that, among the metric theories of gravity individuated by EEP, seem to single-out only those that possess a ``minimal nonlinearity'', in a quite precise sense. Remarkably, this property seems to characterise the purely metric theories, and more specifically those of the Lanczos--Lovelock class (which, in four spacetime dimensions, reduce to Einstein’s gravity).

This conclusion appears to agree with a recent proposal for a formal statement of the SEP.\cite{Gerard:2006ia, Gerard:2008nc} As is well-known, the formalism behind many theories of gravity (including general relativity) can be interpreted in the spirit of gauge theories of fundamental interactions.\cite{gauge} Here, the connection plays the role of the gauge potential, whereas the field strength is represented by the curvature. Within this framework, the ``natural'' form of the field equations is not a statement on the curvature (as in Einstein’s theory) but rather on its first covariant derivative, namely
\begin{equation}
\nabla\!_d {R_{abc}}^d = \kappa\, j_{abc}\;,
\label{gerardeq}
\end{equation}
where ${R_{abc}}^d$ is the Riemann tensor, $\kappa$ is a coupling constant, and $j_{abc}$ is a current playing the role of source for gravity.\cite{fnn}  Gravity then appears as the Yang--Mills field associated with the Lorentz group.  In particular, the non-linearity of the field equations~\eqref{gerardeq} --- hence, the gravitational self-interaction --- is associated with the fact that the Lorentz group is non-Abelian.  In the words of Ref.~\onlinecite{Gerard:2006ia}: ``Gravitons gravitate like gluons glue.''  In vacuum, Eq.~\eqref{gerardeq} reduces to 
\begin{equation}
\nabla\!_d {R_{abc}}^d= 0\;,
\label{eq:gerard}
\end{equation}
which still contains, however, the information about the non-linearity of the gravitational field.

Remarkably, upon implementing Eq.~\eqref{eq:gerard} as a set of constraints on an asymptotically flat, spherically symmetric metric representing a weak, stationary gravitational field, one finds exactly the same conditions that  guarantee the validity of the GWEP (more exactly, of NEP for self-gravitating bodies).\cite{Gerard:2006ia, Gerard:2008nc}  Since the GWEP is one of the ingredients of the SEP, Eq.~\eqref{eq:gerard} was then suggested in Ref.~\onlinecite{Gerard:2006ia} as a necessary condition for the SEP to hold.  Although the soundness of this proposal has still to be confirmed in general, it is tantalising to see again a connection between non-linearity and the SEP. An extension of this idea to other theories of gravity, and to higher dimensions, might thus reveal new unexpected links between the GNEP, the GWEP, and the SEP.

The path to a deep understanding of the ``family tree" of gravitation theories  is probably still long and tortuous, but perhaps the equivalence principles, or their modern offspring, will have a word to say in this story.


\section*{Acknowledgements}

The authors are grateful to two anonymous referees for their constructive criticisms, which led to several improvements in the presentation. E.D.C. acknowledges John C.~Miller, for his constructive criticism, and Vincenzo Vitagliano, for his invaluable support.  S.L. wishes to thank Ian Vega and Matt Visser, for their keen remarks and stimulating questions. S.S. is grateful to Antony Valentini for many friendly fights about the equivalence principle.



\begin{thebibliography}{99}
\bibitem{fn3}
As it is common in classical mechanics, by ``particle'' we mean a body whose extension is irrelevant for what concerns its dynamical behaviour. Moreover, we exclude the possibility that the mentioned particle may carry any spin, quadrupole moment, or any other higher multipole moment whatsoever.
\bibitem{fn4}
For example, $r$ could be defined as the radius of the smallest sphere containing the body, evaluated as $r=(A/4\pi)^{1/2}$, where $A$ is the area of the sphere.
\bibitem{tides}
In practice, for each finite resolution of the measuring device, there is a size below which tidal effects become indistinguishable from noise. This, however, does not make EEP valid --- not even ``for all practical purposes''. The correct statement is that, for experiments performed at a given size in a gravitational field, there is always a sufficiently high resolution that allows one to detect tidal effects. This is, of course, true if spacetime curvature at an event has to be an observable quantity.
\bibitem{noWEP}
Incidentally, this implies that there is no need to include the WEP explicitly in the statement of EEP.  The condition that a gravitational field be locally eliminable (usually implemented through the requirements of local position invariance and local Lorentz invariance) is entirely sufficient.
\bibitem{fn7}
It should also be pointed out, however, that the issue of whether gravity is a manifestation of physics or of geometry is, at this stage, not amenable to experimental verification, and might then not be very relevant.  As repeatedly stressed by Poincar\'e, whether a universal behaviour is due to physics or to geometry is ultimately a matter of convention.  And perhaps it is simply wrong to regard ``geometry'' as something intrinsically different from ``physics'': if one defines ``geometry'' as the ``physics with a universal character'' (but still able to make experimentally verifiable predictions), the subject of the debate becomes just a semantic issue.
\bibitem{fn8}
For example, Nordstr\"om's scalar theory of gravity, which contains only one field equation for a relativistic gravitational potential, is too simple for that.  The equation of motion for test particles in such a theory can --- and must --- be postulated independently, exactly as it happens in Newton's theory.  This can be traced to the fact that this theory contains a non-dynamical object: the Minkowski metric. In scalar theories of gravity, the WEP and the GWEP constrain only the equation of motion, not the field equations.
\bibitem{fnn}
In Einstein's general relativity, $\kappa=8\pi G/c^4$, and 
%
\[j_{abc}=\nabla\!_b T_{ac}-\nabla\!_a T_{bc}+\frac{1}{2}\left(\g_{bc}\nabla\!_a T-\g_{ac}\nabla\!_bT\right),\]
%
where $T_{ab}$ is the stress-energy-momentum tensor, and $T:=\g^{ab}\,T_{ab}$ is its trace.
\end{thebibliography}
\end{document}